\documentclass[11pt]{article}

\usepackage[margin=1in]{geometry}
\usepackage{amsmath,amssymb,amsfonts,amsthm}
\usepackage{xcolor}
\usepackage[pagebackref]{hyperref}
\usepackage[capitalize,nameinlink,noabbrev]{cleveref}

\hypersetup{
    colorlinks,
    linkcolor={red!100!black},
    citecolor={blue!100!black},
    urlcolor={blue!60!black},
}
\newtheorem{theorem}{Theorem}
\newtheorem{lemma}[theorem]{Lemma}
\newtheorem{definition}[theorem]{Definition}

\title{A note on the partition bound for one-way  classical~communication~complexity}

\begin{document}

\author{Srinivasan Arunachalam\thanks{IBM Quantum, \textsf{Srinivasan.Arunachalam@ibm.com}} \and
Joao F. Doriguello 
\thanks{HUN-REN Alfr\'ed R\'enyi Institute of Mathematics, Budapest, Hungary, \textsf{doriguello@renyi.hu}}
 \and
Rahul Jain\thanks{Centre for Quantum Technologies and Department of Computer Science, National University of Singapore, Singapore, \textsf{rahul@comp.nus.edu.sg}}
}

\maketitle
\begin{abstract}
We present a linear program for a one-way version of the partition bound of Jain and Klauck, denoted $\mathsf{prt}^1_\varepsilon(f)$, and show that it characterizes the one-way randomized communication complexity $\mathsf{R}_\varepsilon^1(f)$ with shared randomness and worst-case error $\varepsilon$ of any relation $f\subseteq\mathcal{X}\times\mathcal{Y}\times\mathcal{Z}$. More precisely, for all $\varepsilon,\delta\in(0,1/2)$, $\mathsf{R}_\varepsilon^1(f) \geq \log\mathsf{prt}_\varepsilon^1(f)$ and $\mathsf{R}_{\varepsilon+\delta}^1(f) \leq \lceil\log\mathsf{prt}_\varepsilon^1(f) + \log\log(1/\delta)\rceil$. This improves upon the characterization of $\mathsf{R}_\varepsilon^1(f)$ in terms of the rectangle bound due to Jain, Klauck, and Nayak [STOC'08], which is tight only up to an additive $O(\log(1/\delta))$ term.
\end{abstract}

\section{Introduction}

The two-party communication model was introduced by Yao~\cite{Yao:1979:CQR:800135.804414}. Two players, Alice and Bob, receive inputs $x\in\mathcal{X}$ and $y\in\mathcal{Y}$, respectively, and must produce an output $z\in\mathcal{Z}$ such that $(x,y,z)$ belongs to a given relation $f\subseteq\mathcal{X}\times\mathcal{Y}\times\mathcal{Z}$, while exchanging as few bits as possible. Despite its simplicity, the model has found applications throughout theoretical computer science. One of the main tasks in communication complexity is to prove lower bounds on the communication required by a given problem, and several general methods have been proposed for randomized and quantum protocols. The earliest ones are the fooling set~\cite{Yao:1979:CQR:800135.804414}, (approximate) rank~\cite{mehlhorn1982vegas,krause1996geometric,buhrman2001communication}, discrepancy~\cite{chor1988unbiased,babai1989multiparty}, and rectangle/corruption~\cite{yao1983lower,babai1986complexity,RAZBOROV1992385,kushilevitz97,klauck2003rectangle} bounds.

Since then, several new and stronger lower bounds have been discovered. Jain, Klauck, and Nayak~\cite{jain2008direct} proposed the subdistribution bound, a relaxation of the rectangle/corruption bound. Klauck~\cite{klauck2007lower} introduced the smooth-discrepancy bound, while Linial and Shraibman~\cite{linial2007lower} introduced the factorization norm ($\gamma_2$) bound, which also holds for quantum protocols; Sherstov~\cite{sherstov2008pattern} later showed that the two coincide. All these bounds were subsumed by the partition bound of Jain and Klauck~\cite{jain2010partition}, a linear program whose optimal value lower bounds the randomized communication complexity of any relation; they also introduced relaxed versions of it, the smooth-rectangle and smooth-discrepancy bounds. This left the field with a single unified lower bound method for randomized communication complexity. Laplante, Lerays, and Roland~\cite{laplante2012classical} then introduced the quantum partition bound, also known as the efficiency bound, defined via \emph{zero-communication protocols with abort}, in which the players do not communicate at all but are allowed to abort. The efficiency bound subsumes the factorization norm when shared entanglement is allowed, and coincides with the partition bound when shared randomness is allowed.

In this note we consider the \emph{one-way} setting, in which Alice sends a single message to Bob, who then produces the output. One-way protocols are the simplest nontrivial special case of the model and have been studied extensively in their own right. For $\varepsilon\in(0,1/2)$ and a relation $f$, let $\mathsf{R}^1_\varepsilon(f)$ denote the one-way randomized communication complexity of $f$ with shared randomness and worst-case error at most $\varepsilon$, i.e., the minimum number of bits Alice must send so that Bob's output is correct with probability at least $1-\varepsilon$ on every input $(x,y)$. (All definitions are made precise in \Cref{sec:prt}.) Jain, Klauck, and Nayak~\cite{jain2008direct} showed that the rectangle bound characterizes $\mathsf{R}^1_\varepsilon(f)$ up to an additive $O(\log(1/\delta))$ term, at the cost of increasing the error by $\delta$.

We present a linear program for a one-way version of the partition bound, whose optimal value we denote by $\mathsf{prt}^1_\varepsilon(f)$, and show that it characterizes $\mathsf{R}^1_\varepsilon(f)$ up to an additive $\log\log(1/\delta)$ term, again at the cost of increasing the error by $\delta$. All logarithms in this note are base $2$. Our main result is the following.
\begin{theorem}
    \label{thr:main}
    For $\delta,\varepsilon\in(0,1/2)$, $\mathsf{R}^1_{\varepsilon}(f) \geq \log\mathsf{prt}_{\varepsilon}^1(f)$ and $\mathsf{R}^1_{\varepsilon+\delta}(f) \leq \lceil\log\mathsf{prt}_\varepsilon^1(f) + \log\log(1/\delta)\rceil$. 
\end{theorem}
The proof has two steps, both of which go through a one-way version of the zero-communication protocols with abort mentioned above. In such a protocol Alice, without communicating, decides whether to abort, and Bob outputs some $z\in\mathcal{Z}$; the protocol is required not to abort with the same probability $\eta$ on every input, and to be correct with probability at least $1-\varepsilon$ conditioned on not aborting. The largest achievable $\eta$ is the \emph{one-way efficiency} $\mathsf{eff}^1_\varepsilon(f)$, a worst-case-error version of the one-way efficiency bound of Laplante, Lerays, and Roland~\cite{laplante2012classical}. In the first step (\Cref{thr:thr1}) we show that $\mathsf{prt}^1_\varepsilon(f) = 1/\mathsf{eff}^1_\varepsilon(f)$: feasible solutions to our linear program correspond exactly to one-way zero-communication protocols. In the second step (\Cref{thr:thr2,thr:thr3}) we relate one-way zero-communication protocols to one-way communication protocols. A $c$-bit communication protocol yields a zero-communication protocol with non-abort probability $2^{-c}$, by guessing Alice's message using the shared randomness. Conversely, a zero-communication protocol with non-abort probability $\eta$ yields a communication protocol in which Alice sends Bob the index of the first of $T\approx\log(1/\delta)/\eta$ shared random strings on which she does not abort; encoding this index costs $\lceil\log T\rceil$ bits, which is where the $\log\log(1/\delta)$ term comes from. 

In~\cite{laplante2012classical}, the one-way efficiency bound is formulated as a linear program for simulating a target distribution on both players' outputs, with error handled by smoothing, and the one-way tightness result concerns the quantum version of the bound. For classical one-way protocols,~\cite{laplante2012classical} refers to the rectangle bound of~\cite{jain2008direct}.

\section{The one-way partition bound}
\label{sec:prt}

Consider a relation $f\subseteq\mathcal{X}\times\mathcal{Y}\times\mathcal{Z}$.
In a two-party one-way communication protocol $\mathcal{P}$, Alice is given $x\in\mathcal{X}$ and sends a message to Bob with input $y\in\mathcal{Y}$. We assume that for every $(x,y)\in\mathcal{X}\times\mathcal{Y}$ given as input, there is always at least one $z\in\mathcal{Z}$ such that $(x,y,z)\in f$. Upon receiving Alice's message, Bob produces the output of the protocol $\mathcal{P}(x,y)$. We shall always assume shared randomness $R\in\{0,1\}^\ast$ between the two players. Since private randomness can be included in $R$, we assume without loss of generality that, for a shared random string $r$, Alice's message $M(x,r)$ is a deterministic function of $x$ and $r$, and Bob's output $\mathcal{P}(m,r,y)$ is a deterministic function of Alice's message $m$, $r$, and $y$. The error of the protocol on input $(x,y)\in\mathcal{X}\times\mathcal{Y}$ and its worst-case error are, respectively,
\[
    \operatorname{err}_{x,y}(\mathcal{P},f) := \Pr[(x,y,\mathcal{P}(x,y)) \notin f]
    \qquad\text{and}\qquad
    \operatorname{err}(\mathcal{P},f) := \max_{(x,y)\in \mathcal{X}\times\mathcal{Y}}\operatorname{err}_{x,y}(\mathcal{P},f).
\]
The classical one-way communication complexity of $f$ with worst-case error at most $\varepsilon$ is
\[
    \mathsf{R}^1_{\varepsilon}(f) := \min_{\mathcal{P}}\{\text{bits communicated by}~\mathcal{P}:\operatorname{err}(\mathcal{P},f)\leq \varepsilon\}.
\]
For ease of notation we do not include the superscript ``pub'' to signal that shared randomness is allowed. 

In a one-way zero-communication protocol with abort, where the output set is $\mathcal{Z}\cup\{{\perp}\}$, Alice and Bob are given $x\in\mathcal{X}$ and $y\in\mathcal{Y}$, respectively. Using shared randomness $R\in\{0,1\}^\ast$, but without communicating to each other, Alice outputs $z_a\in\{\top,{\perp}\}$, where ${\perp}$ means that she aborts, and Bob outputs $z_b\in\mathcal{Z}$. As before, we assume that, for a shared random string $r$, these are deterministic functions $z_a(x,r)$ and $z_b(y,r)$. If $z_a = {\perp}$, the protocol's output, denoted by $\mathcal{P}^{\perp}(x,y)$, is ${\perp}$, otherwise it is $z_b$. The error of the protocol on input $(x,y)\in\mathcal{X}\times\mathcal{Y}$ given that it does not abort, and its worst-case error, are, respectively,
\begin{align*}
    \operatorname{err}^{\perp}_{x,y}(\mathcal{P}^{\perp},f) &:= \Pr_R[(x,y,\mathcal{P}^{\perp}(x,y)) \notin f\,|\,\mathcal{P}^{\perp}(x,y)\neq {\perp}]\\
    \text{and}\qquad \operatorname{err}(\mathcal{P}^{\perp},f) &:= \max_{(x,y)\in \mathcal{X}\times\mathcal{Y}}\operatorname{err}^{\perp}_{x,y}(\mathcal{P}^{\perp},f).
\end{align*}
The non-abort probability of $\mathcal{P}^{\perp}$ on input $(x,y)\in\mathcal{X}\times\mathcal{Y}$ is
\[
    \mathsf{eff}^1(\mathcal{P}^{\perp},(x,y)) := \Pr_R[\mathcal{P}^{\perp}(x,y) \neq {\perp}].
\]
We require this probability to be the same for all $(x,y)\in \mathcal{X}\times\mathcal{Y}$, and denote it by $\mathsf{eff}^1(\mathcal{P}^{\perp})$. The one-way zero-communication \emph{efficiency} of $f$ with worst-case error at most $\varepsilon$~\cite{laplante2012classical} is
\[
    \mathsf{eff}^1_{\varepsilon}(f) := \sup_{\mathcal{P}^{\perp}}\{\mathsf{eff}^1(\mathcal{P}^{\perp}):\operatorname{err}(\mathcal{P}^{\perp},f)\leq \varepsilon\}.
\]
We will see in \Cref{thr:thr1} that the supremum is attained.

We now present our linear program for the one-way partition bound.
\begin{definition}[One-way partition bound]
    Given $\varepsilon\in(0,1/2)$, the one-way $\varepsilon$-partition bound of $f$, denoted by $\mathsf{prt}_\varepsilon^1(f)$, is the optimal value of the following linear program.
\par\medskip
\noindent\begin{minipage}[t]{0.47\textwidth}
    \centerline{\bf Primal}
{\footnotesize 
\begin{align*}
    &\min \sum_{A\subseteq\mathcal{X}} w_A,\\
    &\forall x\in\mathcal{X}: \sum_{A\subseteq\mathcal{X} : x\in A} w_A = 1,\\
    &\forall A\subseteq\mathcal{X},y\in\mathcal{Y}: \sum_{z\in\mathcal{Z}} w_{A,y,z} = w_A,\\
    &\forall (x,y)\in \mathcal{X}\times\mathcal{Y}: \!\! \sum_{A\subseteq\mathcal{X}:x\in A} \sum_{z\in\mathcal{Z}:(x,y,z)\in f} \!\! w_{A,y,z} \geq 1- \varepsilon,\\
    &\forall A\subseteq\mathcal{X}, (y,z)\in\mathcal{Y}\times\mathcal{Z}: w_A \geq 0, w_{A,y,z} \geq 0.
\end{align*}
}%
\end{minipage}%
\hfill
\begin{minipage}[t]{0.47\textwidth}
    \centerline{\bf Dual}
{\footnotesize 
\begin{align*}
    &\max \sum_{(x,y)\in \mathcal{X}\times\mathcal{Y}}(1 - \varepsilon)\mu_{x,y} - \sum_{x\in\mathcal{X}}\lambda_x,\\
    &\forall A\subseteq\mathcal{X},(y,z)\in\mathcal{Y}\times\mathcal{Z}: \sum_{x\in A:(x,y,z)\in f} \mu_{x,y} \leq \lambda_{A,y},\\
    &\forall A\subseteq\mathcal{X}: \sum_{y\in\mathcal{Y}} \lambda_{A,y} \leq 1 + \sum_{x\in A}\lambda_x,\\
    &\forall A\subseteq\mathcal{X}, (x,y)\in\mathcal{X}\times\mathcal{Y}: \mu_{x,y} \geq 0, \lambda_{A,y} \geq 0, \lambda_x \in\mathbb{R}.
\end{align*}
}%
\end{minipage}
\end{definition}
In order to prove \Cref{thr:main}, we start by showing that optimal one-way zero-communication protocols are equivalent to optimal solutions to our partition-bound linear~program.
\begin{lemma}
    \label{thr:thr1}
    For all $\varepsilon\in(0,1/2)$, $\mathsf{prt}_\varepsilon^1(f) = 1/\mathsf{eff}^1_\varepsilon(f)$.
\end{lemma}
\begin{proof}
    We first show that $\mathsf{prt}_\varepsilon^1(f) \geq 1/\mathsf{eff}^1_\varepsilon(f)$. Consider an optimal solution for the primal of $\mathsf{prt}_\varepsilon^1(f)$ with weights $w_A$ and $w_{A,y,z}$ for all $A\subseteq\mathcal{X}$, $(y,z)\in\mathcal{Y}\times\mathcal{Z}$, so that $\sum_{A\subseteq\mathcal{X}} w_A = \mathsf{prt}_\varepsilon^1(f)$. We define a one-way zero-communication protocol as follows: using public coins, Alice and Bob sample $A\subseteq\mathcal{X}$ with probability $w_A/\mathsf{prt}_\varepsilon^1(f)$. Alice does not abort if and only if $x\in A$. Bob outputs $z_b=z$ with probability $w_{A,y,z}/w_A$, which is a valid distribution by the second constraint of the primal. For all $x\in\mathcal{X}$, the probability that Alice does not abort is
    \begin{align*}
        \operatorname{Pr}[z_a \neq {\perp}] = \operatorname{Pr}[x\in A] = \frac{1}{\mathsf{prt}_\varepsilon^1(f)}\sum_{A\subseteq\mathcal{X}:x\in A} w_A = \frac{1}{\mathsf{prt}_\varepsilon^1(f)},
    \end{align*}
    using the first constraint of the primal. Moreover, for all $(x,y)\in \mathcal{X}\times\mathcal{Y}$,
    \begin{align*}
        \operatorname{Pr}[(x,y,z_b) \in f|z_a \neq {\perp}] &= \frac{\operatorname{Pr}[(x,y,z_b) \in f,\ x\in A]}{\operatorname{Pr}[x\in A]} = \mathsf{prt}_\varepsilon^1(f)\sum_{A\subseteq\mathcal{X}:x\in A}\frac{w_A}{\mathsf{prt}_\varepsilon^1(f)}\sum_{z\in\mathcal{Z}:(x,y,z)\in f}\frac{w_{A,y,z}}{w_A}\\
        &= \sum_{A\subseteq\mathcal{X}:x\in A} \sum_{z\in\mathcal{Z}:(x,y,z)\in f} w_{A,y,z} \geq 1-\varepsilon,
    \end{align*}
    by the third constraint of the primal (sets $A$ with $w_A = 0$ are never sampled and contribute nothing to the sums). 
    Hence, our one-way zero-communication protocol has non-abort probability $1/\mathsf{prt}^1_\varepsilon(f)$ and worst-case error at most $\varepsilon$. Thus $\mathsf{eff}^1_\varepsilon(f) \geq 1/\mathsf{prt}_\varepsilon^1(f)$.
    
    On the other direction, we now prove that $\mathsf{prt}_\varepsilon^1(f) \leq 1/\mathsf{eff}^1_\varepsilon(f)$. Consider any one-way zero-communication protocol $\mathcal{P}^{\perp}$ with public randomness $R$, worst-case error at most $\varepsilon$, and non-abort probability $\eta$ for all $(x,y)\in \mathcal{X}\times\mathcal{Y}$. We show that $\mathsf{prt}_\varepsilon^1(f) \leq 1/\eta$; taking the supremum over $\mathcal{P}^{\perp}$ then gives the claim. Given a public random string $r$ with probability $p(r)$, let $\mathcal{X}_r := \{x\in\mathcal{X}|z_a(x,r)\neq {\perp}\}$ be the set of inputs for which Alice does not abort. Define the weights
    \begin{align*}
        \forall A\subseteq\mathcal{X}, (y,z)\in\mathcal{Y}\times\mathcal{Z}: \qquad  w'_A := \frac{1}{\eta}\sum_{r:A = \mathcal{X}_r} p(r) \qquad\text{and}\qquad 
        w'_{A,y,z} := \frac{1}{\eta}\sum_{r:A = \mathcal{X}_r, z_b(y,r) = z}p(r).
    \end{align*}
    Since Alice's decision to abort depends only on $x$ and $r$,
    $$\forall x\in\mathcal{X}: \qquad\sum_{A\subseteq\mathcal{X}:x\in A}w_A' = \frac{1}{\eta}\sum_{r:z_a(x,r)\neq{\perp}}p(r) = \frac{1}{\eta}\operatorname{Pr}_R[\mathcal{P}^{\perp}(x,y) \neq {\perp}] = 1.$$
    Also, since Bob always outputs some $z\in\mathcal{Z}$,
    $$\forall A\subseteq\mathcal{X}, y\in\mathcal{Y}: \qquad  \sum_{z\in\mathcal{Z}} w'_{A,y,z} = \frac{1}{\eta}\sum_{r:A = \mathcal{X}_r}p(r) = w'_A.$$
    Finally, since Bob always outputs a correct answer with probability at least $1-\varepsilon$ given that Alice does not abort,
    \begin{align*}
        \forall (x,y)\in \mathcal{X}\times\mathcal{Y}: \quad 1- \varepsilon &\leq \operatorname{Pr}[(x,y,\mathcal{P}^{\perp}(x,y)) \in f|\mathcal{P}^{\perp}(x,y)\neq{\perp}] \\
        &= \frac{\operatorname{Pr}[(x,y,\mathcal{P}^{\perp}(x,y)) \in f,\mathcal{P}^{\perp}(x,y)\neq{\perp}]}{\operatorname{Pr}[\mathcal{P}^{\perp}(x,y)\neq{\perp}]} \\
        &= \frac{1}{\eta} \sum_{r:x\in \mathcal{X}_r,(x,y,z_b(y,r))\in f}  p(r) = \sum_{A\subseteq\mathcal{X}:x\in A} \sum_{z\in\mathcal{Z}:(x,y,z)\in f} w'_{A,y,z}.
    \end{align*}
    Thus, the weights $w'_A$ and $w'_{A,y,z}$ are a feasible solution for the primal of $\mathsf{prt}_\varepsilon^1(f)$ and therefore
    \[
        \mathsf{prt}_\varepsilon^1(f) \leq \sum_{A\subseteq\mathcal{X}} w'_A = \frac{1}{\eta}\sum_r p(r) = \frac{1}{\eta}. \qedhere
    \]
\end{proof}
Note that the first part of the proof constructs a protocol with non-abort probability exactly $1/\mathsf{prt}_\varepsilon^1(f)$ (the linear program is feasible and bounded, so its optimal value is attained), while the second part shows that no protocol does better. Hence the supremum in the definition of $\mathsf{eff}^1_\varepsilon(f)$ is attained.

In the next two lemmas, we relate one-way randomized and one-way zero-communication protocols, i.e., we show that an optimal one-way randomized communication protocol can simulate a one-way zero-communication protocol and vice-versa. We note that related versions of the next result have appeared before in~\cite{massar2002nonlocality,buhrman2003combinatorics,laplante2012classical}.
\begin{lemma}
    \label{thr:thr2}
    For all $\varepsilon\in (0,1/2)$, $\mathsf{R}_{\varepsilon}^1(f) \geq \log(1/\mathsf{eff}_{\varepsilon}^1(f))$.
\end{lemma}
\begin{proof}
    Consider a one-way randomized communication protocol $\mathcal{P}$ with public coin $R$, worst-case error at most $\varepsilon$, and communication complexity $c:=\mathsf{R}^1_\varepsilon(f)$. Without loss of generality, we assume that Alice's message has length exactly $c$ for all inputs and public coins (shorter messages can be padded). Consider the following one-way zero-communication protocol $\mathcal{P}^{\perp}$. Using their public coin, Alice and Bob sample, independently, $m$ uniformly from $\{0,1\}^c$ and $r$ from the distribution $R$. Alice checks if her transcript $M(x,r)$ from protocol $\mathcal{P}$ corresponding to input $x\in\mathcal{X}$ and public coin $r$ equals $m$. If $M(x,r) \neq m$, she aborts. Bob, on the other hand, outputs $\mathcal{P}(m,r,y)$, which is his output in protocol $\mathcal{P}$ with public coin $r$, Alice's message $m$, and his input $y$. For every $x\in\mathcal{X}$ and $r$, $\Pr_m[M(x,r) = m] = 2^{-c}$, hence $\operatorname{Pr}[\mathcal{P}^{\perp}(x,y) \neq {\perp}] = 2^{-c}$ for all $(x,y)\in \mathcal{X}\times\mathcal{Y}$. Moreover, since this probability does not depend on $r$, conditioned on $M(x,r)=m$ the string $r$ is still distributed according to $R$, and $m = M(x,r)$ is exactly Alice's message in $\mathcal{P}$. Therefore Bob's output $\mathcal{P}(m,r,y)$, conditioned on not aborting, is distributed exactly as the output of $\mathcal{P}$ on $(x,y)$, and so
    $$\forall (x,y)\in \mathcal{X}\times\mathcal{Y}: \quad \Pr[(x,y,\mathcal{P}^{\perp}(x,y)) \in f|\mathcal{P}^{\perp}(x,y) \neq {\perp}] = \Pr_R[(x,y,\mathcal{P}(x,y)) \in f] \geq 1-\varepsilon.$$
    We conclude that 
    \[\mathsf{eff}^1_\varepsilon(f) \geq 2^{-c} \implies \mathsf{R}^1_\varepsilon(f) \geq \log(1/\mathsf{eff}^1_\varepsilon(f)). \qedhere\]
\end{proof}

\begin{lemma}
    \label{thr:thr3}
    For all $\delta,\varepsilon\in(0,1/2)$, $\mathsf{R}^1_{\varepsilon + \delta}(f) \leq \lceil\log(1/\mathsf{eff}_{\varepsilon}^1(f)) + \log\log(1/\delta)\rceil$.
\end{lemma}
\begin{proof}
    Consider a one-way zero-communication protocol $\mathcal{P}^{\perp}$ with public coin $R$, worst-case error at most $\varepsilon$, and efficiency $\mathsf{eff}^1_\varepsilon(f)$. Consider the following one-way randomized communication protocol $\mathcal{P}$: Alice and Bob use public coins to obtain a sequence $r_1,\dots,r_T$ of $T$ independent random strings distributed according to $R$. Alice looks at the minimum index $j\in[T]$ such that, on $r_j$, she does not abort in $\mathcal{P}^{\perp}$, and sends $j$ to Bob, who then outputs $\mathcal{P}^{\perp}(x,y)$ using $r_j$. Call an index $j\in[T]$ \emph{good} if Alice does not abort on $r_j$. Since $\operatorname{Pr}_R[\mathcal{P}^{\perp}(x,y) \neq {\perp}] = \mathsf{eff}^1_\varepsilon(f)$ for all $(x,y)\in \mathcal{X}\times\mathcal{Y}$ and the $r_j$ are independent, by choosing $T = 2^{\lceil\log(\log(1/\delta)/\mathsf{eff}^1_\varepsilon(f))\rceil}$, so that $T\geq \log(1/\delta)/\mathsf{eff}^1_\varepsilon(f)$, we get
    \begin{align*}
        \operatorname{Pr}[\text{no good}~j\in[T]] = (1 - \mathsf{eff}^1_\varepsilon(f))^T \leq e^{-\mathsf{eff}^1_\varepsilon(f)\, T} \leq e^{-\log(1/\delta)} \leq \delta.
    \end{align*}
    If there is no good index, Alice sends $j=1$ and Bob proceeds as above (Bob need not know whether $j$ is good; in this case the protocol may err with probability at most $1$). If there is a good index, then, conditioned on $j$ being the first good index, $r_j$ is distributed according to $R$ conditioned on Alice not aborting on $x$ (the event that $r_j$ is good depends only on $r_j$, and the $r_i$ are independent), so Bob's output is correct with probability at least $1-\varepsilon$. Hence, for all $(x,y)\in\mathcal{X}\times\mathcal{Y}$, the error probability of the protocol is
    \begin{align*}
        \operatorname{Pr}[(x,y,\mathcal{P}(x,y))\notin f] &\leq \operatorname{Pr}[(x,y,\mathcal{P}(x,y))\notin f|\text{some good}~j] + \operatorname{Pr}[\text{no good}~j] \\
        &\leq \varepsilon + \delta.
    \end{align*}
   Since $T$ is a power of $2$, the index $j\in[T]$ can be sent using exactly $\log{T} = \lceil\log(1/\mathsf{eff}_\varepsilon^1(f)) + \log\log(1/\delta)\rceil$ bits. 
\end{proof}

Our main result, \Cref{thr:main}, is thus a straightforward corollary from \Cref{thr:thr1,thr:thr2,thr:thr3}.

\paragraph*{Acknowledgements.} This paper was supported by ERC grant No.\ 810115-DYNASNET and the National Research Foundation, Singapore and A*STAR under the CQT Bridging Grant and the Quantum Engineering Programme Award number NRF2021-QEP2-02-P05.

\bibliographystyle{alphadoi}
\bibliography{bibliography}

\end{document}